\documentstyle[aps,graphics,epsfig,prd,manuscript]{revtex}
\begin{document}
\rightline{\small{\today}}
\vspace{5mm}
\begin{center}

{\bf A Comment on the Roe-Woodroofe Construction of Poisson Confidence
Intervals }

\vspace{5mm}

Mark Mandelkern and Jonas Schultz\\
Department of Physics and Astronomy\\
University of California, Irvine, California 92697\\
(submitted to Phys. Rev. D on Feb. 17, 2000)
\vspace{5mm}
\end{center}

\begin{abstract}

We consider the Roe-Woodroofe construction of confidence intervals for the
case of a Poisson distributed variate where the mean is the sum of a known
background and an unknown non-negative signal. We point out that the
intervals do not have coverage in the usual sense but can be made to have
such with a modification that does not affect the believability and other
desirable features of this attractive construction. A similar modification
can be used to provide coverage to the construction recently proposed by
Cousins for the Gaussian-with-boundary problem. 

\vspace{5mm} 
\noindent PACS number(s): 02.50.Cw, 02.50.Kd, 02.50.Ph, 06.20.Dk
\end{abstract}

\newpage
\section{Introduction}
\hspace{5mm}

A problem of long-standing interest is that of setting confidence
intervals for an unknown non-negative signal $\mu$ in the presence of a
known mean background $b$ when the measurement $n$ is Poisson distributed
as $p(n;\mu+b)$. When $n<b$, the usual estimate for $\mu$, i.e. $n-b$, is
negative, leading in most constructions to small upper limits that imply
unrealistically high confidence in small values of $\mu$. In a recent
paper, Roe and Woodroofe \cite{bib:rw} propose a construction that
produces more believable intervals and contains the unifying feature that
one need not decide beforehand whether to set a confidence interval or an
upper confidence bound. However, since the Roe-Woodroofe confidence belt
(of confidence level $\alpha$) is not constructed from an unconditional
probability density and does not have coverage in the usual sense (i.e.
unconditional coverage), one cannot state that the unconditional
probability of the interval enclosing the true value is at least $\alpha$.
Our comment is that a straightforward modification of the Roe-Woodroofe
confidence belt gives it coverage, making the construction effectively an
{\it ordering principle} applied to the Poisson pdf, albeit reached by
circuitous means.

\section{Roe-Woodroofe Confidence Intervals}
\hspace{5mm}

Roe and Woodroofe are motivated by the observation \cite{bib:rw} that the
measurement $n=0$ implies that zero signal (as well as zero background) is
seen; thus, the resulting estimate for $\mu$ is zero, independent of $b$. 
They argue therefore that the confidence interval for $\mu$ for n=0 must
be independent of $b$. Extending the argument, they note that for any
observation $n$, one has observed a result $n$ from the Poisson pdf
$p(n;\mu+b)$ {\it and} a background of at most $n$. They formulate a
method of obtaining confidence intervals based on the conditional
probability to observe $n$ given a background $\leq n$ and obtain the
desired result for $n=0$ and approximately the classical confidence
intervals for $n>b$.  While they identify their method as an ordering
principle, it is not one in the same sense as Ref.s \cite{bib:fc} and
\cite{bib:giunti} which explicitly choose a confidence belt of probability
$\alpha$ using the Poisson pdf $p(n;\mu+b)$ and the Likelihood Ratio
Construction and invert it to find confidence intervals. The latter
methods do not obtain intervals that are independent of $b$ for $n=0$ and
yield confidence intervals which are unphysically small for $n<b$. 

\begin{figure}
\scalebox{0.8}[0.8]{\includegraphics[1cm,3cm][22cm,22cm]{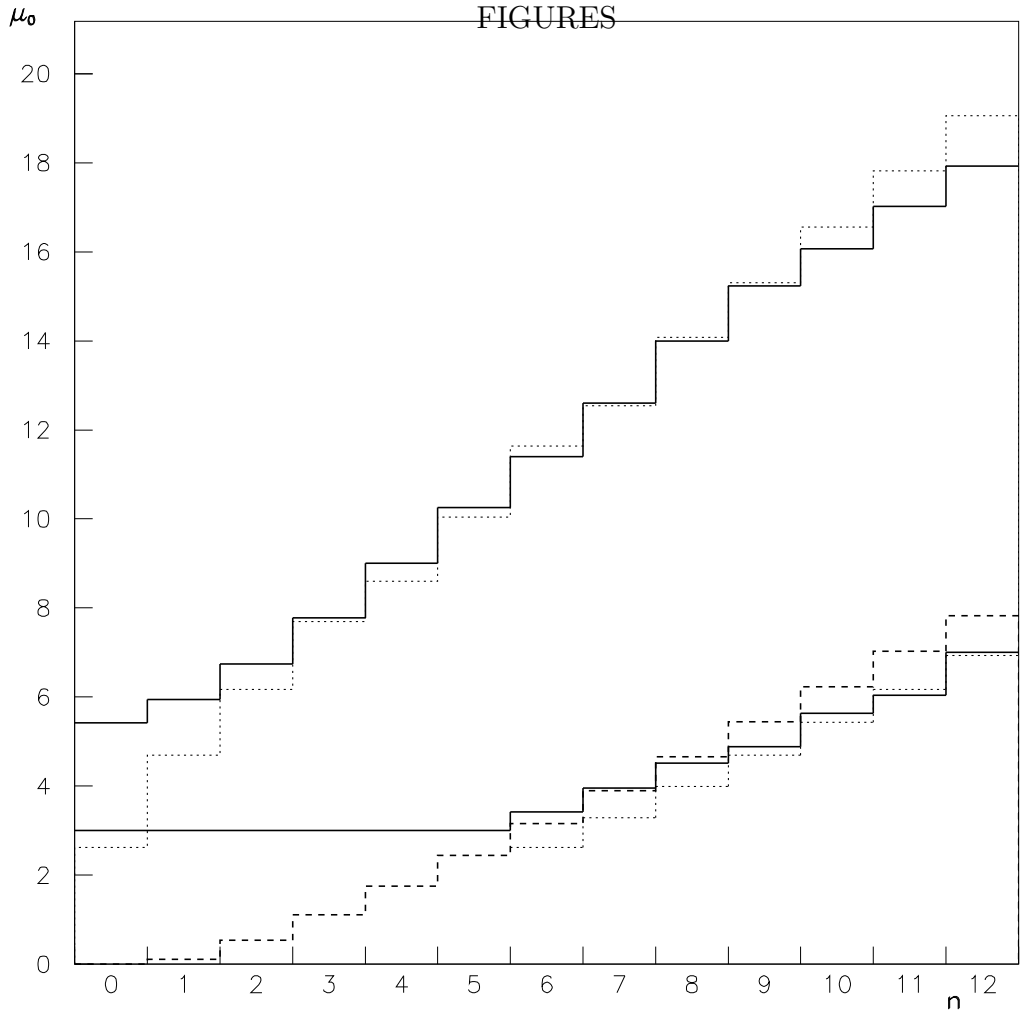}}
\caption{ 90\% Poisson confidence belts for unknown non-negative
signal $\mu$ in the presence of a background with known mean b
taken to be 3.0, where $n$ is the result of a single observation. The
solid belt is the Roe-Woodroofe construction, the dotted belt the
central construction and the dashed belt the one-sided construction of
90\% Poisson lower limits. Here $\mu_0=\mu+b$ is the parameter
representing the mean of signal plus background. We illustrate confidence 
belts in this manner to demonstrate the absence of coverage for the
Roe-Woodroofe construction and to emphasize that a naive approach to
setting a confidence interval for $\mu$ leads to a null interval for
sufficiently small $n<b$, in this case $n=0$. The solid line Roe-Woodroofe
lower limit for $n \leq 5$ is at $\mu=0$.}
\label{fig:poisson_rw} 
\end{figure}

\begin{figure}
\scalebox{0.90}[0.90]{\includegraphics[1cm,0.5cm][22cm,22cm]
{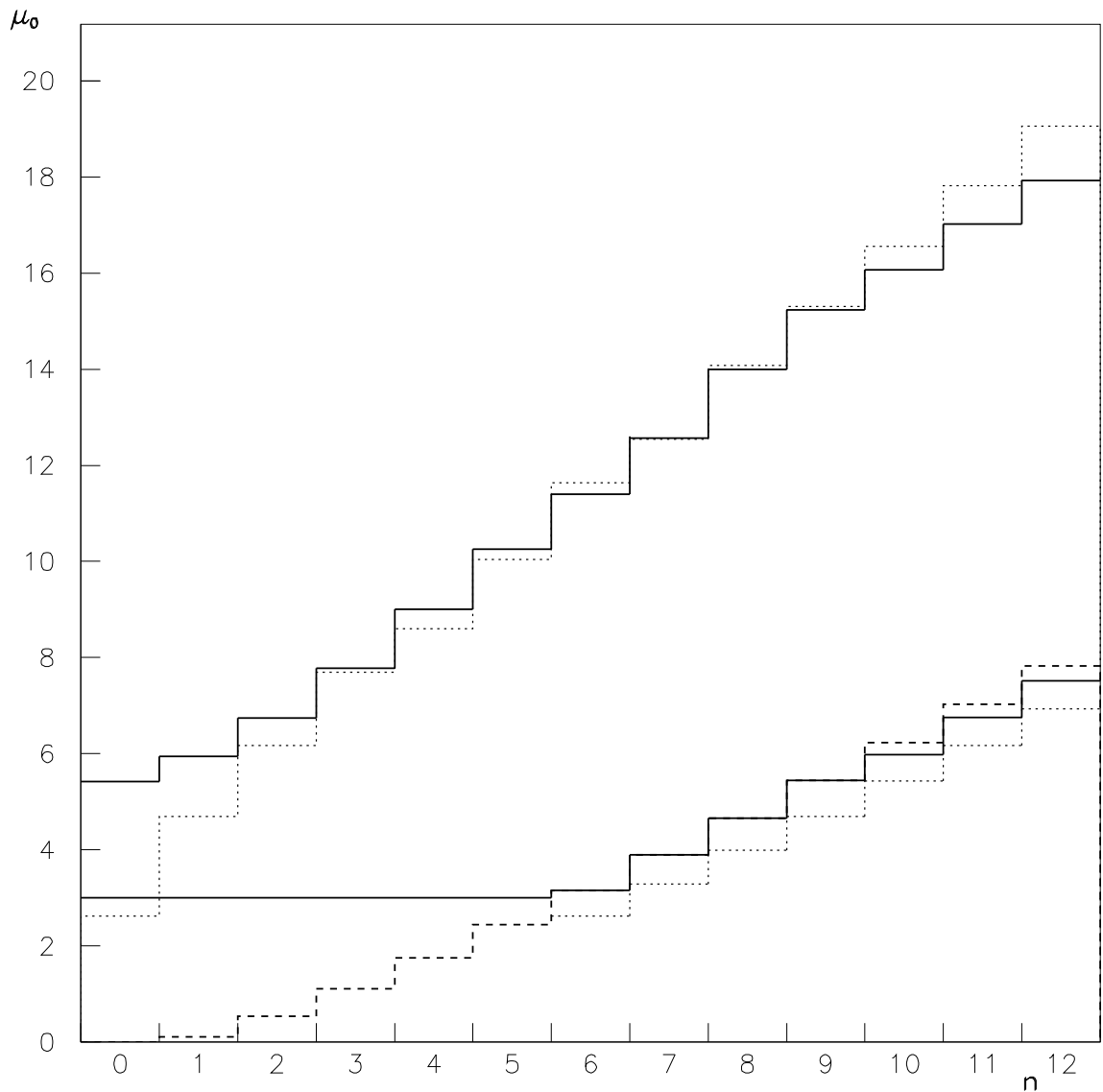}}
\caption{90\% Poisson confidence belts described in Fig.
\ref{fig:poisson_rw}
where the solid belt is modified as described in the text to give
coverage. The dotted and dashed belts are described in the Fig.
\ref{fig:poisson_rw} caption. For $n=6,7,8,9$ the lower limits of the
confidence  intervals coincide with the one-sided 90\% Poisson lower
limits. This guarantees $\geq$90\% probability within the horizontal
intervals.} 
\label{fig:poisson_rwcov} 
\end{figure}

Although the Roe-Woodroofe construction does not have coverage in the
usual sense , it can be easily modified to obtain coverage, by retaining
the left-hand boundary of the confidence belt and adjusting the right-hand
boundary so that for all $\mu$ the horizontal intervals contain
probability $\alpha$.  In Fig. \ref{fig:poisson_rw} we show the
Roe-Woodroofe 90\% intervals for $b=3$ along with one-sided and central
confidence belts \footnote{We show the confidence belt consisting of
central intervals [$n_1(\mu_0)$, $n_2(\mu_0)$] containing at least 90\% of
the probability for unknown Poisson mean $\mu_0$ in the absence of any
known background (dotted)  and the 90\% one-sided belt consisting of
intervals [0, $n_{os}(\mu_0)$](dashed).  There is some arbitrariness in
the choice of a central interval for a discrete variate. We choose the
smallest interval such that there is $\geq$ 90\% of the probability in the
center and $\leq$5\%, but as close as possible to 5\%, on the right. The
alternative of requiring $\leq$5\%, but as close as possible to 5\%, on
the left gives slightly less symmetrical intervals. For the latter choice
the 90\% Poisson upper limit for $n=0$ is $\mu_0=3.0$ compared to
$\mu_0=2.62$ for our choice. For $\mu_0<2.62$, according to this
prescription, one cannot construct an interval containing probability
$>90\%$ that does not include $n=0$ and we adopt 90\% one-sided
intervals.} for the Poisson distribution without background. We note that
the Roe-Woodroofe horizontal intervals do not coincide with the one-sided
intervals shown for $\mu < 2.44$. Therefore for some values of $\mu$ in
this range, the confidence belt does not satisfy the coverage requirement
that $\geq$90\% of the probability is contained.  Because coverage cannot
be exact when the variable is discrete, the {\it error} for the example
given here is not of great numerical significance. The minimum coverage of
$\sim 0.87$ is obtained at $\mu\sim0.4$. Undercoverage is more severe for
greater $b$; for $b=10.0$, the minimum coverage is $\sim 0.78$. However,
it is desirable to have coverage, which we obtain as shown in Fig.
\ref{fig:poisson_rwcov} where we have changed the right side of the
confidence belt so that the horizontal intervals contain probability
$\geq$90\%. We note that the confidence intervals for small $n$, i.e. 
$n<b$, are unchanged. Intervals for both constructions are given in Table
\ref{table:rw}.

It would be nice to devise an ordering principle that can be directly
applied to the Poisson pdf $p(n;\mu+b)$ to obtain the confidence belt
shown in Fig. \ref{fig:poisson_rwcov}, if only because the construction we
have used here is aesthetically unpleasing. This method, which consists of
first determining vertical intervals per Ref. \cite{bib:rw}, and then
fixing them, leaves something to be desired. However, in the end the
method of construction does not really matter. What results here is an
ordering procedure that yields a confidence belt with coverage and
produces physically sensible intervals.

B. Roe has noted \cite{bib:roe} that our modification is equally
applicable to a construction due to R. Cousins, in which the Roe-Woodroofe
method of conditioning is applied to the Gaussian-with-boundary
\cite{bib:cousins} problem. Here, for example, an interval of confidence
level $\alpha$ is sought for an unknown non-negative signal $\mu$ and the
measurements x are normally distributed as N(x;$\mu$). As for the
Roe-Woodroofe construction referred to above, the Cousins construction
produces physically sensible confidence intervals for all $x$ including $x
< 0$. However this construction significantly undercovers for $\mu < 0.5$
and significantly overcovers for $\mu \sim 1$. In order to produce exact
coverage using the Cousins construction, we retain the left hand (upper)
curve of the confidence belt $x_l(\mu)$ and recalculate the right hand
(lower) curve $x_r(\mu)$ so that the horizontal intervals contain
probability $\alpha$ using: 

\begin{equation} 
2 \alpha = erf(\frac{\mu-x_l}{\sqrt 2})+erf(\frac{x_r-\mu}{\sqrt 2}).
\end{equation}

\section{Conclusion}
\hspace{5mm}

For the case of Poisson distributed measurements $n$ with a non-negative
signal mean $\mu$ and known mean background $b$, the Roe-Woodroofe
construction produces well-behaved confidence intervals, particularly for
$n<b$ where other constructions yield unphysically small intervals. Since
the construction is not based on integrating probabilities that arise from
an unconditional pdf, it does not produce a confidence belt with coverage
in the usual frequentist sense. We suggest a modification that provides
coverage while preserving the desirable features of the construction. 
While the changes introduced by this modification are relatively small for
the example given here (they are larger for greater b), nevertheless the
procedure corrects a formal defect in the original construction. A similar
modification provides coverage for a construction recently discussed by R.
Cousins for the Gaussian-with-boundary problem.

\begin{table}
\begin{centering}
\begin{tabular}{|c|c|c|c|c|}
\hline
 & \multicolumn{2}{c|}{Roe-Woodroofe}&\multicolumn{2}{c|}{Modified}\\
$n$(observed) & Lower & Upper & Lower & Upper\\ \hline
0    & 0.0 & 2.44 & 0.0 & 2.44 \\   
1    & 0.0 & 2.95 & 0.0 & 2.95 \\
2    & 0.0 & 3.75 & 0.0 & 3.75 \\
3    & 0.0 & 4.80 & 0.0 & 4.80 \\
4    & 0.0 & 6.01 & 0.0 & 6.01 \\
5    & 0.0 & 7.28 & 0.0 & 7.28 \\
6    & 0.42 & 8.40 & 0.16 & 8.42 \\
7    & 0.96 & 9.58 & 0.90 & 9.58 \\
8    & 1.52 & 10.99 & 1.66 & 11.02 \\
9    & 1.88 & 12.23 & 2.44 & 12.23 \\
10   & 2.64 & 13.50 & 2.98 & 13.51 \\
11   & 3.04 & 14.80 & 3.75 & 14.77 \\
12   & 4.01 & 15.90 & 4.52 & 16.01 \\
\end{tabular}
\caption{ Comparison of confidence intervals for the Roe-Woodroofe and
modified Roe-Woodroofe constructions}
\label{table:rw}
\end{centering}
\end{table}

\end{document}